\documentclass[pra,twocolumn]{revtex4}

\usepackage{amssymb, amsmath}
\usepackage[dvips]{graphicx}
\usepackage{hyperref}
\usepackage{epsf}
\usepackage{pstricks, pst-node, pst-plot}

\newcommand{\ket}[1]{| #1 \rangle}

\newcommand{\inprod}[2]{\langle #1 | #2 \rangle}

\begin{document}
\title{Numerical evidence for the maximum number of mutually unbiased bases in dimension six}
\author{Paul Butterley}
\email{pb504@york.ac.uk}
\author{William Hall}
\email{wah500@york.ac.uk}
\affiliation{Department of Mathematics, University of York, Heslington, York YO10 5DD, UK}

\begin{abstract}
The question of determining the maximal number of mutually unbiased bases in dimension six has received much attention since their introduction to quantum information theory, but a definitive answer has still not been found. In this paper we move away from the traditional analytic approach and use a numerical approach to attempt to determine this number. We numerically minimise a non-negative function $\mathfrak{N}_{d,N}$ of a set of $N+1$ orthonormal bases in dimension $d$ which only evaluates to zero if the bases are mutually unbiased. As a result we find strong evidence that (as has been conjectured elsewhere) there are no more than three mutually unbiased bases in dimension six.
\end{abstract}


\maketitle

\section{Introduction}

Mutually unbiased bases have uses in a variety of topics in quantum mechanics.  The first person to consider their use was Ivanovic \cite{JPhysA.14.3241} who introduced them in the problem of state determination.  Here it is found that measurements based on mutually unbiased bases are optimal for constructing the state of an ensemble of systems.  This application was investigated further by Wootters and Fields some time later \cite{AoP.191.363}.  In 2002 Cerf, Bourennane, Karlsson and Gisin \cite{PhysRevLett.88.127902} explicitly use mutually unbiased bases to extend the BB84 \cite{BB84} and the six-state protocol \cite{PhysRevLett.81.3018} quantum cryptographic schemes \footnote{The qutrit case was considered at roughly the same time by Bechmann-Pasquinucci and Peres \cite{PhysRevLett.85.3313} and Bru\ss{} and Macchiavello \cite{PhysRevLett.88.127901}.}.  The \emph{Mean King's Problem} is also related to mutually unbiased bases \cite{PRA.284.1,PRA.71.052331,PRA.73.050301}.

Mutually unbiased bases are simple to define. Suppose we have a $d$-dimensional Hilbert space and a number of orthonormal bases for this space described by vectors $\ket{\psi_{k,m}}$, where $m$ labels one of the vectors in basis $k$.  We call these bases \emph{mutually unbiased} if
\begin{equation}
	\left|\inprod{\psi_{k,m}}{\psi_{l,n}}\right| = \frac{1}{\sqrt{d}} \qquad \forall k \neq l, \quad \forall m,n = 1,\ldots,d.
\end{equation}
A commonly asked question is then: What is the maximum number of mutually unbiased bases that can be found for general $d$?

Ivanovic, in his paper on state determination \cite{JPhysA.14.3241}, constructed a set of $d+1$ mutually unbiased bases for all prime $d$. Wootters and Fields \cite{AoP.191.363} went further and proved not only that $d+1$ is the maximal possible number of mutually unbiased bases in dimension $d$, but that this bound is realised for all prime power dimensions. 

The smallest dimension for which the maximum number of mutually unbiased bases is unknown is $d=6$. It is known \cite{Finite.Fields.and.Applications} that for any dimension a the maximal number of mutually unbiased bases $\mathfrak{M}(d)$ can be bounded below e.g.
\begin{align*}
\mathfrak{M}(d) &\geq \min(\mathfrak{M}(p_1^{e_1}), \ldots, \mathfrak{M}(p_r^{e_r})) \\
&= \min(p_1^{e_1}+1,\ldots,p_r^{e_r}+1),
\end{align*}
where $d=p_1^{e_1}\ldots p_r^{e_r}$ is the prime factorisation of $d$.  Thus it is clear that $\mathfrak{M}(6) \geq 3$.  A number of the constructions for three mutually unbiased bases in $d=6$ have been examined by Grassl \cite{qph.0406175} and it has been proven that these cannot be extended further.  Archer \cite{JMathPhys.46.022106} also proved that the methods used to determine $\mathfrak{M}(d)$ for prime power dimensions cannot be generalised to non-prime power dimensions. Bengtsson et al. \cite{qph.0610161} consider a constructions from a number of known types of Hadamard matrices but are unable to find more than three mutually unbiased bases for $d=6$ from these. Other results establishing an improved lower bound for square dimensions exist \cite{qph.0407081}, but the general question is far from answered, and the above inequality is the only known bound for $d=6$.

The inability to analytically find four or more mutually unbiased bases in dimension six has brought about a general belief that $\mathfrak{M}(6) = 3$ (as conjectured in \cite{Zauner}).  Rather than approach this problem analytically, we will reformulate it in such a way that we can perform a numerical minimisation to provide evidence to support or disprove this conjecture.  This approach has been suggested by Bengtsson \cite{qph.0610161}, and a similar calculation has been carried out independently in \cite{qph.0612035}; however, to our knowledge, no literature presenting detailed results of a numerical investigation exists.

The rest of this article is organised as follows.  In section 2 we discuss our reformulation of the problem into a minimisation problem and the algorithm we use to find the minima.  In section 3 we present the results of our various searches and analyse them.  Finally in section 4 we discuss other potential approaches to the problem and summarise our results.  We also provide more technical details of the minimisation algorithm in the appendix.

\section{Numerical methods for mutually unbiased bases}

In this section we show how the problem of finding mutually unbiased bases can be cast in the form of an optimisation problem. We will illustrate how this problem is constructed and how it can be attacked numerically.

\subsection{Constructing the optimisation problem}

Let us first state the general idea we are trying to pursue here. We wish to create a function of $N+1$ orthonormal bases (if they existed) that would achieve a global minimum when these $N+1$ bases are mutually unbiased. We then hope to obtain the minimum using an appropriate optimisation algorithm.

We start with some useful preliminaries. Let $\ket{\phi_m} = \sum_{n=1}^d \phi_{mn} \ket{n}$ be an orthonormal basis for $m=1, \ldots, d$. Define a $d \times d$ matrix $U$ by the coefficients of the individual vectors i.e. $U_{nm} = \phi_{mn}$ (so that column $m$ corresponds to $\ket{\phi_m}$); then the fact that the basis is orthonormal implies that $U$ is unitary.  Now suppose that $A, B$ are two unitary matrices representing two orthonormal bases $\ket{\alpha_m} = \sum_{n=1}^d A_{nm} \ket{n}$ and $\ket{\beta_m} = \sum_{n=1}^d B_{nm} \ket{n}$ respectively. Then 
\begin{equation} \left(A^\dagger B\right)_{mn} = \sum_{p=1}^d A^*_{pm}B_{pn} = \inprod{\alpha_m}{\beta_n} \end{equation}
and so if the two bases are mutually unbiased, then $\left|\left(A^\dagger B\right)_{mn}\right|= 1/\sqrt{d}$ for all $m,n=1,\ldots,d$.

So, we can represent $N+1$ mutually unbiased bases by unitary matrices $U_1, \ldots, U_N, U_{N+1} \equiv \openone$ (where without loss of generality we have rotated every basis so that the last basis is simply the standard basis $\{ \ket{k} \}_{k=1}^d$). So now we are looking for a function of the matrix elements $\left|\left(U_k^\dagger U_l\right)_{mn}\right|$ that is minimised when each of these norms is equal to $1/\sqrt{d}$. One such function is
\begin{align}
& \mathfrak{N}_{d,N}(U_1, \ldots, U_N) = \notag \\
& \qquad \sum_{1 \leq k < l \leq N+1} \sum_{m,n=1}^d \left( \left|\left(U_k^\dagger U_l\right)_{mn}\right|^2 - \frac{1}{d} \right)^2 \label{mub_norm_min}
\end{align}
which has a minimum of zero when all of the matrices represent a set of $N$ mutually unbiased bases \footnote{This function when evaluated for a pair of orthogonal basis can be thought of as a measure of how close the two bases are to being mutually unbiased. The function also has an interpretation as a `distance' function; see \cite{qph.0610161} for details.}.  So in principle we can determine whether a set of $N+1$ mutually unbiased bases exists by finding the \emph{global} minimum of this function over all sets of unitaries $U_k$. Using the fact that we can write any unitary matrix $U$ in the form $U=e^{iH}$, where $H$ is Hermitian, we can state this minimisation in an \emph{unconstrained} form i.e.
\begin{equation} \label{mub_min_problem}
\textrm{Minimise } \mathfrak{N}_{d,N}(e^{iH_1}, \ldots, e^{iH_N})
\end{equation}

\subsection{Implementing the minimisation numerically}

The function $\mathfrak{N}(U_1, \ldots, U_N)$ has the form of a sum of squares of a series of functions, and so the minimisation problem stated in (\ref{mub_min_problem}) is a non-linear least squares optimisation problem. This kind of optimisation problem can be solved using the Levenberg-Marquadt algorithm \cite{QuartApplMath.2.164,SIAM.11.431,More.1977}. However, like many such algorithms, it is only guaranteed to find a local rather than a global maximum, so to try and resolve this issue we will run the algorithm many times from a randomised starting point.

The computational mathematics package MATLAB has an implementation of both this algorithm and a fast and accurate matrix exponentiation routine, and so we choose to use this package to implement our algorithm. Some of the more technical details of the implementation are given in the Appendix.

\section{Results}
Our tests for $d=6$ concentrated on both finding a full set of $d+1$ mutually unbiased bases, which exist in prime power dimensions, and also finding four MUBs which is a less computationally complex problem.  For comparison we also conducted a number of tests for $d=2,3,4,5,7$.  In all tests we consider a set of mutually unbiased bases to be found if a set of unitary matrices are found such that $\mathfrak{N}_{d,N}(U_1, \ldots, U_N) \leq 10^{-6}$.  The results are presented in Figure \ref{fig:testresults}.

Our attempts to find four mutually unbiased bases for $d=6$ provide us with strong evidence towards their non-existence.  Although the success rate was not high, our minimisation method is able to find four mutually unbiased bases for $d=4,5,7$.  The method was unable to find four mutually unbiased bases in dimension six and Figure \ref{fig:results63} presents the frequency of minima obtained.  The peaks suggest that in general our algorithm is converging to a small number of local minima.  Moreover, more than two-thirds of the test runs converged to the value 0.051249, the minimum value of $\mathfrak{N}_{6,3}$ our algorithm was able to find.

For $d<6$ the success rate of finding $d+1$ mutually unbiased bases was very high.  When attempting to find seven mutually unbiased bases for dimension six we obtain a similar picture as when trying to obtain four.  In this case just under a third of the runs converged to our minimum value $1.584472$. The long run time for minimising $\mathfrak{N}_{7,7}$ \footnote{On a 2.8GHz Pentium 4 with 512Mb of memory, each individual optimisation takes at least 10 minutes.} prevents us from obtaining a larger number of results for $d=7$, but on a small number of occasions, a full set of mutually unbiased bases is found. On inspecting the distribution of the obtained results, we do not observe the same grouping of the non-zero minima as in Figure \ref{fig:results63}, suggesting that the algorithm may not be converging to a local minimum in most cases.

\begin{center}
\begin{figure*}[!tp]
\begin{tabular}{|c|c|c|c|c|c|c|}
\cline{2-7}
\multicolumn{1}{c}{} & \multicolumn{3}{|c}{$d+1$ MUBs} & \multicolumn{3}{|c|}{$4$ MUBs}\\ 
\hline
Dimension $d$ & No. of tests & $d+1$ MUBs found & \% success & No. of tests & $4$ MUBs found & \% success \\
\hline 
2 & 2500 & 2500 & 100\% & N/A & N/A & N/A \\
3 & 2500 & 2499 & 99.9\%    & N/A & N/A & N/A \\
4 & 2500 & 2500 & 100\% & 2500 & 2500 & 100\% \\ 
5 & 2500 & 2495 & 99.8\% & 2500 & 1510 & 60.4\% \\
6 & 2500 & 0 & 0\% & 10000 & 0 & 0 \% \\
7 & 250  & 3  & 1.2\% & 3000 & 26 & 0.9 \% \\
\hline
\end{tabular}
\caption{Results illustrating number of times each minimisation problem converges to zero, i.e. a set of mutually unbiased bases is obtained.}
\label{fig:testresults}
\end{figure*}
\end{center}

\begin{center}
\begin{figure}[!bp]

\begin{pspicture}(-1,-1)(10,8)
\psset{yunit=0.0008cm}
\psset{xunit=16cm}
\psaxes[Dx=0.05, Dy=1000](0,0)(0.451,8000) 

\psframe[fillcolor=black,fillstyle=solid,linewidth=0.5pt](0.05,0)(0.055,7070)
\psframe[fillcolor=black,fillstyle=solid,linewidth=0.5pt](0.175,0)(0.18,386)
\psframe[fillcolor=black,fillstyle=solid,linewidth=0.5pt](0.195,0)(0.2,983)
\psframe[fillcolor=black,fillstyle=solid,linewidth=0.5pt](0.21,0)(0.215,131)
\psframe[fillcolor=black,fillstyle=solid,linewidth=0.5pt](0.23,0)(0.235,994)
\psframe[fillcolor=black,fillstyle=solid,linewidth=0.5pt](0.26,0)(0.265,121)
\psframe[fillcolor=black,fillstyle=solid,linewidth=0.5pt](0.265,0)(0.27,96)
\psframe[fillcolor=black,fillstyle=solid,linewidth=0.5pt](0.275,0)(0.28,111)
\psframe[fillcolor=black,fillstyle=solid,linewidth=0.5pt](0.29,0)(0.295,68)
\psframe[fillcolor=black,fillstyle=solid,linewidth=0.5pt](0.325,0)(0.33,13)
\psframe[fillcolor=black,fillstyle=solid,linewidth=0.5pt](0.335,0)(0.34,7)
\psframe[fillcolor=black,fillstyle=solid,linewidth=0.5pt](0.355,0)(0.36,1)
\psframe[fillcolor=black,fillstyle=solid,linewidth=0.5pt](0.405,0)(0.41,6)
\psframe[fillcolor=black,fillstyle=solid,linewidth=0.5pt](0.425,0)(0.43,5)
\psframe[fillcolor=black,fillstyle=solid,linewidth=0.5pt](0.435,0)(0.44,1)

\rput[b]{90}(-0.07,4000){Frequency}
\rput[t]{0}(0.225,-800){Minimal value}

\end{pspicture}

\caption{\label{fig:results63} Histogram of numerically determined minimal values of $\mathfrak{N}_{6,3}$. The separation of the histogram bars indicates that we really are obtaining local minima from our calculations.} \end{figure}
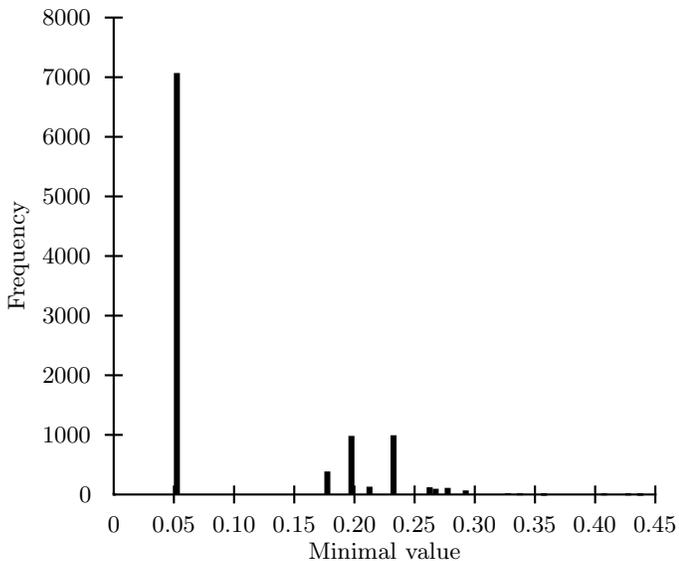

\end{center}

\section{Conclusion}

The numerical results we have presented here represent clear evidence that a set of more than three mutually unbiased bases do not exist in dimension six. It is also clear from the results in Figure \ref{fig:testresults} that this approach does not scale well at all as the dimension of the bases increases. However, we have presented only one possible approach to this problem. Our choice of minimising function is by no means unique. Another plausible test function could be based on the Shannon entropy $H(x_1,\ldots,x_d) = -\sum_i x_i \log x_i$, which, when $\sum_i x_i = 1$, is maximised when $x_i=1/d$. We would of course have to use a different optimisation routine in this case. Another possibility is to simply parameterise the unitary matrices $U_i$ elementwise, and impose the unitarity condition as a constraint on the elements of each of the matrices. We have considered these approaches briefly, and found for the calculations we considered that our method was fastest. We also chose to approximate first derivatives via finite differences rather than calculate them directly; again, this was faster, and appeared to have no effect on accuracy, but this may not be the case with other methods. However, given that we are working with a function of $O(d^2N)$ variables, minimising $\mathfrak{N}_{d,N}$ (or indeed any other appropriate function) efficiently and effectively may well become very difficult using any algorithm as $d$ and $N$ increase \footnote{For example, $\mathfrak{N}_{d,N}$ is a function of $k=d^2(N+1)$ variables, and a sum of $l=\frac{1}{2}d^2N(N+1)$ square terms. The time complexity of the Levenberg-Marquardt algorithm is $O(l^3)$ \cite{NumOpt,DM}, which for this example is $O(d^6N^6)$ i.e. we have quick growth with respect to both $d$ and $N$.}. An example of a different method (although presented in less detail) can be found in \cite{qph.0612035} (see also \footnote{The bases found by the method used in \cite{qph.0612035} can also be used as winning strategies for the challenged party in the mean king's problem (see paper for details).}).

We can also consider the more general question of how close a set of $N+1$ orthonormal bases can be to being mutually unbiased.
The minimisation function used here is simply one way of quantifying how close a particular set of orthonormal bases is to being mutually unbiased (albeit one that has a nice geometric interpretation \cite{AoP.191.363,qph.0610216}). However, there are clearly other ways of doing this. One possibility is to relax the actual definition of mutually unbiased e.g. we could define two vectors to be approximately mutually unbiased if $-\epsilon_+ \leq |\inprod{\psi}{\phi}| - 1/\sqrt{d} \leq \epsilon_+$ for some $\epsilon_\pm > 0$. In principle, one could determine what $\epsilon_\pm$ have to be to allow the existence of $N+1$ approximately mutually unbiased bases $\ket{\psi_{k,m}}$ by considering the minimum and maximum value of $\left|\inprod{\psi_{k,m}}{\psi_{l,n}}\right| -1/d$ over all sets of $N+1$ orthonormal bases. However, since these functions are not differentiable at many points, there may well be difficulties in finding the local extrema of these minimum and maximum functions (our brief attempt at trying this numerically yielded problems even for 3 bases in dimension 2). Knowing how close a set of orthonormal bases can get to being mutually unbiased could be helpful for example in determining efficient state determination protocols, as the information gained from a set of measurements in these bases simply depends on the geometrical relationships between the bases \cite{AoP.191.363}.

\acknowledgments{The authors would like to thank the participants of the York QIS for inspiration leading to this work, and Tony Sudbery for his ongoing support and reading this manuscript. We would also like to thank Michael Reimpell and Reinhard Werner for bringing \cite{qph.0612035} to our attention. PB would like to thank the University of York for a research studentship, and WH would like to thank the Engineering and Physical Sciences Research Council for research funding.}

\appendix

\section{Technical details of implementation}

Our implementation as stated uses the MATLAB implementation of the Levenberg-Marquadt non-linear least squares minimisation routine (\texttt{lsqnonlin}). For speed, the evaluation of the function $\mathfrak{N}_{d,N}(U_1, \ldots, U_N)$ was coded in C, and compiled within MATLAB. We set the routine to terminate when the value of the function $\mathfrak{N}_{d,N}(U_1, \ldots, U_N)$ changes by a value less than $10^{-8}$ between iterations (this threshold was chosen to give a decent balance between accuracy and time efficiency).

The Levenberg-Marquadt algorithm only allows optimisation over the reals, and so to deal with this we must parameterise the Hermitian matrices $H_k$ in (\ref{mub_min_problem}) using real variables. The C function that evaluates $\mathfrak{N}_{d,N}(U_1, \ldots, U_N)$ then reconstructs the unitaries $U_k$ and evaluates the real valued norms $\left|\left(U_k^\dagger U_l\right)_{mn}\right|$, which are then passed back to the optimisation routine.

As mentioned in the main body of the text, we pick a random starting point for each run of the algorithm. To generate the initial points, we must generate a number of random unitaries $U_k$, and then take the matrix logarithm of these unitaries to determine the values of $H_k$. We generate random unitaries in the Haar measure (the analog of the uniform distribution for compact matrix groups, which can be implemented using the QR factorization. The following MATLAB code generates a random unitary matrix $U$ in the Haar measure \cite{AN.233}:
\begin{center} 
\begin{verbatim}
[Q,R] = qr(randn(dim) + i*randn(dim));
U = Q*diag(exp(2*pi*i*randn(dim,1)));
\end{verbatim} 
\end{center}
To exponentiate and take logarithms of matrices we utilise the in-built MATLAB functions.

\bibliography{mub}

\begin{thebibliography}{24}
\expandafter\ifx\csname natexlab\endcsname\relax\def\natexlab#1{#1}\fi
\expandafter\ifx\csname bibnamefont\endcsname\relax
  \def\bibnamefont#1{#1}\fi
\expandafter\ifx\csname bibfnamefont\endcsname\relax
  \def\bibfnamefont#1{#1}\fi
\expandafter\ifx\csname citenamefont\endcsname\relax
  \def\citenamefont#1{#1}\fi
\expandafter\ifx\csname url\endcsname\relax
  \def\url#1{\texttt{#1}}\fi
\expandafter\ifx\csname urlprefix\endcsname\relax\def\urlprefix{URL }\fi
\providecommand{\bibinfo}[2]{#2}
\providecommand{\eprint}[2][]{\url{#2}}

\bibitem[{\citenamefont{Ivanovic}(1981)}]{JPhysA.14.3241}
\bibinfo{author}{\bibfnamefont{I.~D.} \bibnamefont{Ivanovic}},
  \bibinfo{journal}{Journal of Physics A} \textbf{\bibinfo{volume}{14}},
  \bibinfo{pages}{3241} (\bibinfo{year}{1981}).

\bibitem[{\citenamefont{Wootters and Fields}(1989)}]{AoP.191.363}
\bibinfo{author}{\bibfnamefont{W.~K.} \bibnamefont{Wootters}} \bibnamefont{and}
  \bibinfo{author}{\bibfnamefont{B.~D.} \bibnamefont{Fields}},
  \bibinfo{journal}{Annals of Physics} \textbf{\bibinfo{volume}{191}},
  \bibinfo{pages}{363} (\bibinfo{year}{1989}).

\bibitem[{\citenamefont{Cerf et~al.}(2002)\citenamefont{Cerf, Bourennane,
  Karlsson, and Gisin}}]{PhysRevLett.88.127902}
\bibinfo{author}{\bibfnamefont{N.~J.} \bibnamefont{Cerf}},
  \bibinfo{author}{\bibfnamefont{M.}~\bibnamefont{Bourennane}},
  \bibinfo{author}{\bibfnamefont{A.}~\bibnamefont{Karlsson}}, \bibnamefont{and}
  \bibinfo{author}{\bibfnamefont{N.}~\bibnamefont{Gisin}},
  \bibinfo{journal}{Phys. Rev. Lett.} \textbf{\bibinfo{volume}{88}},
  \bibinfo{pages}{127902} (\bibinfo{year}{2002}).

\bibitem[{\citenamefont{Bennett and Brassard}(1984)}]{BB84}
\bibinfo{author}{\bibfnamefont{C.}~\bibnamefont{Bennett}} \bibnamefont{and}
  \bibinfo{author}{\bibfnamefont{G.}~\bibnamefont{Brassard}}, in
  \emph{\bibinfo{booktitle}{Proceedings of IEEE International Conference on
  Computer, Systems and Signal Processing}} (\bibinfo{year}{1984}), pp.
  \bibinfo{pages}{175--179}.

\bibitem[{\citenamefont{Bru\ss{}}(1998)}]{PhysRevLett.81.3018}
\bibinfo{author}{\bibfnamefont{D.}~\bibnamefont{Bru\ss{}}},
  \bibinfo{journal}{Phys. Rev. Lett.} \textbf{\bibinfo{volume}{81}},
  \bibinfo{pages}{3018} (\bibinfo{year}{1998}).

\bibitem[{\citenamefont{Englert and Aharonov}(2001)}]{PRA.284.1}
\bibinfo{author}{\bibfnamefont{B.-G.} \bibnamefont{Englert}} \bibnamefont{and}
  \bibinfo{author}{\bibfnamefont{Y.}~\bibnamefont{Aharonov}},
  \bibinfo{journal}{Physics Letters A} \textbf{\bibinfo{volume}{284}},
  \bibinfo{pages}{1} (\bibinfo{year}{2001}).

\bibitem[{\citenamefont{Hayashi et~al.}(2005)\citenamefont{Hayashi, Horibe, and
  Hashimoto}}]{PRA.71.052331}
\bibinfo{author}{\bibfnamefont{A.}~\bibnamefont{Hayashi}},
  \bibinfo{author}{\bibfnamefont{M.}~\bibnamefont{Horibe}}, \bibnamefont{and}
  \bibinfo{author}{\bibfnamefont{T.}~\bibnamefont{Hashimoto}},
  \bibinfo{journal}{Physical Review A} \textbf{\bibinfo{volume}{71}},
  \bibinfo{pages}{052331} (\bibinfo{year}{2005}).

\bibitem[{\citenamefont{Kimura et~al.}(2006)\citenamefont{Kimura, Tanaka, and
  Ozawa}}]{PRA.73.050301}
\bibinfo{author}{\bibfnamefont{G.}~\bibnamefont{Kimura}},
  \bibinfo{author}{\bibfnamefont{H.}~\bibnamefont{Tanaka}}, \bibnamefont{and}
  \bibinfo{author}{\bibfnamefont{M.}~\bibnamefont{Ozawa}},
  \bibinfo{journal}{Physical Review A} \textbf{\bibinfo{volume}{73}},
  \bibinfo{pages}{050301(R)} (\bibinfo{year}{2006}), \eprint{quant-ph/0604096}.

\bibitem[{\citenamefont{Klappenecker and
  R{\"o}tteler}(2004)}]{Finite.Fields.and.Applications}
\bibinfo{author}{\bibfnamefont{A.}~\bibnamefont{Klappenecker}}
  \bibnamefont{and}
  \bibinfo{author}{\bibfnamefont{M.}~\bibnamefont{R{\"o}tteler}},
  \emph{\bibinfo{title}{Constructions of Mutually Unbiased Bases}}
  (\bibinfo{year}{2004}).

\bibitem[{\citenamefont{Grassl}(2004)}]{qph.0406175}
\bibinfo{author}{\bibfnamefont{M.}~\bibnamefont{Grassl}}
  (\bibinfo{year}{2004}), \eprint{quant-ph/0406175}.

\bibitem[{\citenamefont{Archer}(2005)}]{JMathPhys.46.022106}
\bibinfo{author}{\bibfnamefont{C.}~\bibnamefont{Archer}},
  \bibinfo{journal}{Journal of Mathematical Physics}
  \textbf{\bibinfo{volume}{46}}, \bibinfo{pages}{022106}
  (\bibinfo{year}{2005}), \eprint{quant-ph/0312204}.

\bibitem[{\citenamefont{Bengtsson et~al.}(2006)\citenamefont{Bengtsson, Bruzda,
  Ericsson, Larsson, Tadej, and Zyczkowski}}]{qph.0610161}
\bibinfo{author}{\bibfnamefont{I.}~\bibnamefont{Bengtsson}},
  \bibinfo{author}{\bibfnamefont{W.}~\bibnamefont{Bruzda}},
  \bibinfo{author}{\bibfnamefont{A.}~\bibnamefont{Ericsson}},
  \bibinfo{author}{\bibfnamefont{J.-A.} \bibnamefont{Larsson}},
  \bibinfo{author}{\bibfnamefont{W.}~\bibnamefont{Tadej}}, \bibnamefont{and}
  \bibinfo{author}{\bibfnamefont{K.}~\bibnamefont{Zyczkowski}}
  (\bibinfo{year}{2006}), \eprint{quant-ph/0610161}.

\bibitem[{\citenamefont{Wocjan and Beth}(2004)}]{qph.0407081}
\bibinfo{author}{\bibfnamefont{P.}~\bibnamefont{Wocjan}} \bibnamefont{and}
  \bibinfo{author}{\bibfnamefont{T.}~\bibnamefont{Beth}}
  (\bibinfo{year}{2004}), \eprint{quant-ph/0407081}.

\bibitem[{\citenamefont{Zauner}(1999)}]{Zauner}
\bibinfo{author}{\bibfnamefont{G.}~\bibnamefont{Zauner}}, Ph.D. thesis,
  \bibinfo{school}{Universit{\"a}t Wien} (\bibinfo{year}{1999}).

\bibitem[{\citenamefont{Reimpell and Werner}(2006)}]{qph.0612035}
\bibinfo{author}{\bibfnamefont{M.}~\bibnamefont{Reimpell}} \bibnamefont{and}
  \bibinfo{author}{\bibfnamefont{R.~F.} \bibnamefont{Werner}}
  (\bibinfo{year}{2006}), \eprint{quant-ph/0612035}.

\bibitem[{\citenamefont{Levenberg}(1944)}]{QuartApplMath.2.164}
\bibinfo{author}{\bibfnamefont{K.}~\bibnamefont{Levenberg}},
  \bibinfo{journal}{Quart. Appl. Math} \textbf{\bibinfo{volume}{2}},
  \bibinfo{pages}{164} (\bibinfo{year}{1944}).

\bibitem[{\citenamefont{Marquardt}(1963)}]{SIAM.11.431}
\bibinfo{author}{\bibfnamefont{D.~W.} \bibnamefont{Marquardt}},
  \bibinfo{journal}{Journal of the Society for Industrial and Applied
  Mathematics} \textbf{\bibinfo{volume}{11}}, \bibinfo{pages}{431}
  (\bibinfo{year}{1963}), ISSN \bibinfo{issn}{0368-4245}.

\bibitem[{\citenamefont{Mor{\'e}}(1977)}]{More.1977}
\bibinfo{author}{\bibfnamefont{J.~J.} \bibnamefont{Mor{\'e}}}, in
  \emph{\bibinfo{booktitle}{Numerical Analysis}}, edited by
  \bibinfo{editor}{\bibfnamefont{G.~A.} \bibnamefont{Watson}}
  (\bibinfo{publisher}{Springer Verlag}, \bibinfo{year}{1977}), Lecture Notes
  in Mathematics 630, pp. \bibinfo{pages}{105--117}.

\bibitem[{\citenamefont{Bengtsson}(2006)}]{qph.0610216}
\bibinfo{author}{\bibfnamefont{I.}~\bibnamefont{Bengtsson}}
  (\bibinfo{year}{2006}), \eprint{quant-ph/0610216}.

\bibitem[{\citenamefont{Edelman and Rao}(2005)}]{AN.233}
\bibinfo{author}{\bibfnamefont{A.}~\bibnamefont{Edelman}} \bibnamefont{and}
  \bibinfo{author}{\bibfnamefont{N.~R.} \bibnamefont{Rao}},
  \bibinfo{journal}{Acta Numerica} p. \bibinfo{pages}{233}
  (\bibinfo{year}{2005}).

\bibitem[{\citenamefont{Bechmann-Pasquinucci and
  Peres}(2000)}]{PhysRevLett.85.3313}
\bibinfo{author}{\bibfnamefont{H.}~\bibnamefont{Bechmann-Pasquinucci}}
  \bibnamefont{and} \bibinfo{author}{\bibfnamefont{A.}~\bibnamefont{Peres}},
  \bibinfo{journal}{Phys. Rev. Lett.} \textbf{\bibinfo{volume}{85}},
  \bibinfo{pages}{3313} (\bibinfo{year}{2000}).

\bibitem[{\citenamefont{Bru\ss{} and
  Macchiavello}(2002)}]{PhysRevLett.88.127901}
\bibinfo{author}{\bibfnamefont{D.}~\bibnamefont{Bru\ss{}}} \bibnamefont{and}
  \bibinfo{author}{\bibfnamefont{C.}~\bibnamefont{Macchiavello}},
  \bibinfo{journal}{Phys. Rev. Lett.} \textbf{\bibinfo{volume}{88}},
  \bibinfo{pages}{127901} (\bibinfo{year}{2002}).

\bibitem[{\citenamefont{Nocedal and Wright}(2000)}]{NumOpt}
\bibinfo{author}{\bibfnamefont{J.}~\bibnamefont{Nocedal}} \bibnamefont{and}
  \bibinfo{author}{\bibfnamefont{S.~J.} \bibnamefont{Wright}},
  \emph{\bibinfo{title}{Numerical Optimization}}
  (\bibinfo{publisher}{Springer}, \bibinfo{year}{2000}).

\bibitem[{\citenamefont{Ye}(2003)}]{DM}
\bibinfo{author}{\bibfnamefont{N.}~\bibnamefont{Ye}}, \emph{\bibinfo{title}{The
  handbook of data mining}} (\bibinfo{publisher}{Lawrence Erlbaum Associates},
  \bibinfo{year}{2003}).

\end{thebibliography}
\end{document}